\documentstyle[aps,prd,psfig]{revtex}
\begin{document}
\title{Can a Very Low-Luminosity and Cold White Dwarf be a \\
Self-gravitating Bose Condensed System}
\author{Nandini Nag$^{a)}$\thanks{E-mail:nandini@klyuniv.ernet.in}
and Somenath Chakrabarty$^{a,b)}$\thanks{E-mail:somenath@klyuniv.ernet.in}\\
 a) Department of Physics, University of Kalyani, Kalyani 741 235,
India\thanks{Permanent address}\\
b) Inter-University Centre for Astronomy and Astrophysics, Post Bag 4\\
Ganeshkhind, Pune 411 007, India}

\date{\today}
\maketitle
\begin{flushleft}
\end{flushleft}

\begin{abstract}
An entirely new model for the structure as well as for the cooling mechanism
of white dwarfs has been proposed. We have argued that the massive part of 
the constituents of white dwarfs- the positively charged ions are boson and 
under the extreme physical condition (density and temperature) at the interior, 
it is possible to have condensation of charged bose gas. We have tried to 
establish that a cold white dwarf is a self gravitating charged bose condensed 
system.
\end{abstract}

Unlike the other branches of physical science, astrophysics deals with
giant macroscopic objects and also tiny constituents of microscopic world.
In the astrophysical calculations very often we use the numbers as large as
Planck mass ($M_p$) and very small number like Planck's constant ($h$).
The observational data from most of the giant macroscopic stellar
objects, e.g., stars, white dwarfs, neutron stars, etc.  are analyzed by
various models based on micro-physical processes. To be more precise,
the processes like
stellar evolutions, supernova explosions, structure of white dwarfs and
neutron stars etc. are explained by microscopic models based on nuclear
physics, hydrodynamics, quantum mechanics, statistical mechanics, etc. 
Various astrophysical processes, e.g., emission of high energy
radiations (in particular X-rays  and $\gamma$-rays), neutrinos, many
physical phenomena from pulsar observations, viz.,  pulsar
glitches, existence of strong magnetic fields in neutron stars, etc.
are also explained  by various micro-physical processes. The theoretical
analysis of thermal
evolution and also the evolution of magnetic fields of neutron stars are
based on the physics of superfluidity, superconductivity etc. of neutron
matter \cite{sap}. Similarly, the theoretical studies on the structure and the  
thermal evolution (which is again related to 
luminosity function) of white dwarfs of different masses and temperatures, 
in which we are especially interested in this article are also 
based on some interesting micro physical models \cite{sal1}. It is believed 
that the structure of white dwarfs is built on the following two
assumptions. Electrons are all degenerate and they contribute to the total
pressure of the system. 
On the other hand, ions are non-degenerate and almost the entire mass of the 
star comes from their rest masses. Based on these two assumptions and 
by employing the condition of hydrostatic equilibrium, the equation for the 
structure of white dwarfs can be obtained. Now the basic constituents of 
white dwarfs are mainly oxygen, carbon, helium with an envelope of hydrogen
gas. It is however, sometime assumed that the whole star is made up of either 
carbon or helium or a composite structure of these three elements, with the
heaviest at the core and lightest at the crust. The structure of a white 
dwarf is strictly determined by its mass. In some relatively massive white 
dwarfs, one can think of the presence of heavier elements like neon, magnesium 
or even iron within the stars.
The class of white dwarfs which have originated  as main sequence stars
with masses around $5M_\odot$ or below are carbon-oxygen ($CO$) white
dwarfs, the other category, which originated from main sequence stars
with masses from $5M_\odot$ to about $12M_\odot$ are
oxygen-neon-magnesium ($ONeMg$) white dwarfs. It is also possible to
have white dwarfs with helium rich interiors. Stars of low enough mass
are supposed to evolve, after its hydrogen burning phase, to a stage
where degeneracy appears before helium burning. But such isolated
stellar objects are not expected to have yet evolved beyond the main
sequence. However, it is possible to have white dwarfs with He-core \cite{apa}
from more massive progenitor in a time scale smaller than the age of the
galaxy if they belong to a binary system which allows mass loss before
He-burning. Now because of high density all these
elements within the white dwarfs are in completely ionized condition. The 
stability of the star against gravitational collapse is supported by 
degenerate electron pressure.  Now the conventional model of white dwarf
cooling is based on the following three consecutive processes: (i) relaxation 
of lattice thermal vibration, (ii) crystallization of the ionic part which are 
in the gaseous/liquid phase and (iii) further release of 
energy by the separation of different masses during crystallization in 
presence of gravity \cite{moc,hug} (see also \cite{rag,maz,ise}). The total 
available energy in these processes depends on
the constituent mass or equivalently on the mass of the white dwarf 
and temperature of the system. 
The time scale for cooling process is $\sim 10$Gyr \cite{sal2}. In this article 
we shall try to establish an entirely new picture for both the structure and
the cooling of white dwarfs based on the condensation of charged bose gas. 
Now it is well known 
from the nuclear structure studies, that the kind of isotopes of 
He, C, O, Mg, Ne, etc. present in white dwarfs are all bosons. 
The white dwarf at relatively high temperature 
is therefore a self gravitating (one-component or multi-component) positively 
charged giant macroscopic bose  system with a back ground of electron gas 
(negatively charged degenerate Fermi system). Based on this fact, we have tried 
to establish from a very simple physical argument that within the 
white dwarf it is possible to achieve a temperature which leads to condensation 
of charged bose gas. We shall also try to explain the cooling of white 
dwarfs with this new physical picture by assuming the condensation 
of bose gas as a first order phase transition.  

Before we go into the detail discussion of bose condensation 
in a cold white dwarf, we give a very brief introduction on the condensation of 
ideal and weakly interacting hard-core bose gases. The condensation of weakly 
interacting 
bose gas- one of the oldest subjects has got a new dimension after the 
remarkable experimental achievement of bose condensation of alkali atoms in 
magnetic traps in the laboratory (\cite{andr}, also see \cite{chr} for an 
overall 
knowledge in this subject). This discovery has given re-birth to this old 
subject and created an enormous interest in both theoretical and experimental 
studies of weakly interacting bose gas \cite{chr,dal}. It is well known that 
in the case of an 
ideal bose gas, the condition determining the fugacity $z$ in the normal phase 
is $ \lambda^3n=g_{3/2}(z)$, where  $n$ is the particle density \cite{hua1}, 
\begin{equation}
\lambda=\left (\frac{2\pi\hbar^2}{mkT}\right )^{1/2}
\end{equation} 
is the thermal wavelength and $g_\nu(z)=\sum_{l=0}^\infty z^\nu/l^\nu$.
Now the condition that the zero momentum state (i.e. the condensed state) is
macroscopically occupied is $\lambda^3n > \zeta(3/2)$, where
$\zeta(3/2)=g_{3/2}(z=1) \simeq 2.612$, the standard zeta function.
Hence the transition temperature for the condensation of ideal bose gas is
given by,
  \begin{equation}
  T_0= \frac{2\pi\hbar^2}{mk} \left [\zeta \left (\frac{3}{2} \right )
  \right ]^{-\frac{2}{3}} n^{\frac{2}{3}}
 \end{equation}
On the other hand for an imperfect bose gas with non negligible hard core 
radius of the constituent, it was shown in several interesting theoretical
calculations in the last few years, that the difference $\bigtriangleup 
T=T_c-T_0 \neq 0$, where $T_c$ is the
critical temperature in the case of imperfect bose gas. But there is no
unique relation for the variation of $\bigtriangleup T$ with scattering length
$a$. Not even there is consensus on the sign of $\bigtriangleup T$. Following
the argument that a spatial repulsion of the constituents is equivalent to the 
momentum space attraction- which leads to condensation of hard core bose
gas at relatively high temperature.
Now for a system of identical bosons, assuming that the interaction 
potential acts locally (i.e. the range of interaction is small compared to
the interparticle distance) and characterized entirely by $s$-wave
scattering length $a$, the limit for which the quantum many body perturbation
is valid is $a \ll \lambda$.
With these assumptions it was shown \cite{hua2,hua3} by Stoof that 
$\bigtriangleup T \sim a^{\frac{3}{2}}$, Bijlsma and Stoof showed that
$\bigtriangleup T \sim a^{\frac{1}{2}}$. The Monte-Carlo simulation gives
$\bigtriangleup T=T_0 C_0(na^3)^{\gamma}$. Where $C_0=0.34 \pm 0.06$
and $\gamma=0.34\pm 0.003$. The mean field calculation shows
$\bigtriangleup T=0.7T_0 (na^3)^{\frac{1}{3}}$. The simple calculation
by Huang shows that $\bigtriangleup T=3.527T_0 (na^3)^{\frac{1}{6}}$.
All these interesting results show that although there is no consensus on how
$\bigtriangleup T$ depends on scattering length, however, there is no
confusion that the condensation of imperfect bose gases occur at relatively 
higher temperature compared to that of an ideal bose gas.

We shall now go back to the discussion on the possibility of bose
condensation in cold white dwarfs. The system we have considered 
contains massive ions carrying $Z$-unit positive charge ($Z$ is fixed 
for one-component systems, on the other hand it is a variable in the case of 
multicomponent systems) which form a non-degenerate charged bose gas
neutralized by a back ground of degenerate electron gas. Now in gases
with short range forces the weak coupling limit corresponds to low
density range, whereas the weak coupling limit in the case of charged
bose gases interacting via coulomb force, this corresponds to the high
density limit. Hence in the case of white dwarfs of mass density
$\geq 10^6$gm/cc, and at temperature $\sim 10^6-10^8$K, 
the weak coupling limit is a good approximation. Therefore, if we
simulate the screened coulomb potential of a positive ion by
some equivalent hard-core potential with range $r_s \approx a$, the
scattering length, we can apply the results of hard-core bose gas in this
dense bosonic system. The scattering length $a$ defined above indicates
the beginning of the repulsive region of the ion.

To get an idea of the order of magnitude of transition temperature in
such dense bosonic systems we equate the thermal de Broglie wavelength 
(eqn.(1)) with
the inter-ionic spacing $\sim n^{-1/3}$. The transition temperature
$T_c^{(1)}$ in $^oK$ obtained from this equality is displayed in Table I
for three different elements (He, C and O). We have considered four
possible mass
densities. The transition temperatures so obtained for various physical 
conditions are found to be well within the limit of 
expected internal temperature of white dwarfs. Therefore the possibility of
bose condensation inside cold white dwarfs can not be ruled out. The
critical temperature $T_c^{(0)}$ obtained from eqn.(2), assuming 
ideal bose gas model is also shown in the table for the same set of
elements and densities. We have also displayed the modified value of
critical temperature $T_c^{(h)}$ from Huang's paper \cite{hua2}
(the values do not
change significantly with other results as cited before). Now it is almost
impossible to obtain the scattering length $a$ within the white dwarf matter, 
we therefore treat it as a fixed parameter. Since we would like to use the 
results derived for hard-core bose gas obtained from perturbative calculation,
keeping in mind the limit of its validity, 
we use the value $a=0.01\lambda$. We believe that even under extreme condition
($\rho = 10^9$gm/cc), the scattering length can not go below this small
value. The thermal wavelengths for two different cases: with and without
hard-core radius are also presented in the table in $\AA$ unit. We have
denoted these two quantities by $\lambda_T^{(h)}$ and $\lambda_T^{(0)}
$ respectively. Assuming that the bose
condensation is a first order phase transition, we have calculated the
total amount of energy  released in the process. The total available energy
is given by
$\bigtriangleup E=L M$, where $L\approx kT_c$, the latent heat of
condensation and $M$ is the total mass of the system. Assuming
$5.0\times 10^3$Km as the radius of a typical white dwarf which 
undergoes a bose condensation transition at a certain critical temperature (for
hard-core gas),
we have shown the amount of energy liberated during this process within the
star for the same set of elements and mass densities as discussed above.
The detail analysis of thermal evolution of white dwarfs taking bose
condensation into account will be presented in some future publication.

From the tabulated results it may be inferred that the critical
temperature for condensation increases with  density. The transition
process will therefore be favored in massive white dwarfs. It is
also obvious that for a given density the lighter ions condense at
relatively higher temperature. Now the ionic distribution in a
multicomponent white dwarf is such that the matter at the core region is 
mainly dominated by
massive ions, whereas the lighter ions built the crustal part. Therefore,
the argument given by Lamb \cite{lam} long ago on the possibility of separation 
of normal and bose condensed phases in presence of gravity may be ruled
out. However, the process strongly depends on so many factors, e.g., the
mass, central density and temperature, the basic constituents and
finally on the radial distribution of mass and temperature  within the star.
If there is a phase separation in presence of gravity, some
extra energy will also be released in the process. A detail analysis will be
presented in a future paper. Therefore the  final conclusion is  that
the possibility of bose condensation can not be ruled out, particularly in
massive white dwarfs of very low luminosity, at least at the central
region. The structure of such massive white dwarfs could therefore be a bose
condensed core with crystalline normal crustal matter. We can compare such
structures with the typical structure  of a neutron star with superfluid
core and normal neutron matter crust. In a future publication, we shall
discuss the most important issue- the controversy of white dwarf ages
in globular clusters using this new ideas of white dwarf structure and
cooling mechanism introduced in this letter.
%

\begin{tabular} {|l|cccc|r|} \hline 
{\em Element}     & 
 \multicolumn{4}{r|}{\em Helium} \\  \hline
         $\rho$ (gm/cc) &$10^6$ &$10^7$ &$10^8$ &$10^9$ \\ \hline
         $T_c^{(1)}$$^o$K&$6.8\times 10^5$&$3.1\times 10^6$&$1.5\times 10^7$
         &$6.8\times 10^7$\\ \hline
         $T_c^{(0)}$$^o$K&$1.1\times 10^5$&$5.3\times 10^5$&$2.4\times 10^6$
         &$1.1\times 10^7$\\ \hline
         $T_c^{(h)}$$^o$K&$2.6\times 10^5$&$1.2\times 10^6$&$5.7\times 10^6$
         &$2.6\times 10^7$\\ \hline
         $\lambda_T^{(0)}$$\AA$&$2.6\times 10^{-2}$&$1.2\times 10^{-2}$&$5.6\times 10^{-3}$
         &$2.6\times 10^{-3}$\\ \hline
         $\lambda_T^{(h)}$$\AA$&$1.7\times 10^{-2}$&$7.9\times 10^{-3}$&$3.7\times 10^{-3}$
         &$1.7\times 10^{-3}$\\ \hline
         $\Delta E$ (ergs)&$1.0\times 10^{46}$&$5.0\times 10^{47}$&$2.0\times 10^{49}$
         &$1.0\times 10^{51}$\\ \hline
{\em Element}     & 
 \multicolumn{4}{r|}{\em Carbon} \\  \hline
         $\rho$ (gm/cc) &$10^6$ &$10^7$ &$10^8$ &$10^9$ \\ \hline
         $T_c^{(1)}$$^o$K&$1.1\times 10^5$&$5.0\times 10^5$&$2.3\times 10^6$
         &$1.1\times 10^7$\\ \hline
         $T_c^{(0)}$$^o$K&$1.8\times 10^4$&$8.5\times 10^4$&$3.9\times 10^5$
         &$1.8\times 10^6$\\ \hline
         $T_c^{(h)}$$^o$K&$4.2\times 10^4$&$2.0\times 10^5$&$9.1\times 10^5$
         &$4.2\times 10^6$\\ \hline
         $\lambda_T^{(0)}$$\AA$&$3.7\times 10^{-2}$&$1.7\times 10^{-2}$&$8.0\times 10^{-3}$
         &$3.7\times 10^{-3}$\\ \hline
         $\lambda_T^{(h)}$$\AA$&$2.5\times 10^{-2}$&$1.1\times 10^{-2}$&$5.3\times 10^{-3}$
         &$2.5\times 10^{-3}$\\ \hline
         $\Delta E$ (ergs)&$6.0\times 10^{44}$&$2.8\times 10^{46}$&$1.7\times 10^{48}$
         &$6.1\times 10^{49}$\\ \hline
{\em Element}     & 
 \multicolumn{4}{r|}{\em Oxygen} \\  \hline
         $\rho$ (gm/cc) &$10^6$ &$10^7$ &$10^8$ &$10^9$ \\ \hline
         $T_c^{(1)}$$^o$K&$6.7\times 10^4$&$3.1\times 10^5$&$1.4\times 10^6$
         &$6.7\times 10^6$\\ \hline
         $T_c^{(0)}$$^o$K&$1.1\times 10^4$&$5.2\times 10^4$&$2.4\times 10^5$
         &$1.1\times 10^6$\\ \hline
         $T_c^{(h)}$$^o$K&$2.6\times 10^4$&$1.2\times 10^5$&$5.6\times 10^5$
         &$2.6\times 10^6$\\ \hline
         $\lambda_T^{(0)}$$\AA$&$4.1\times 10^{-2}$&$1.9\times 10^{-2}$&$8.9\times 10^{-3}$
         &$4.1\times 10^{-3}$\\ \hline
         $\lambda_T^{(h)}$$\AA$&$2.7\times 10^{-2}$&$1.3\times 10^{-3}$&$5.8\times 10^{-3}$
         &$2.8\times 10^{-3}$\\ \hline
         $\Delta E$ (ergs)&$3.0\times 10^{44}$&$1.3\times 10^{46}$&$6.0\times 10^{47}$
         &$2.8\times 10^{49}$\\ \hline
\end{tabular}
\begin{center}
TABLE~~ I
\end{center}
\noindent ACKNOWLEDGEMENT: The work is supported by The Department of
Science \& Technology, Government of India, sanction no.:
SP/S2/K3/97(PRU).
\end{document}